\documentclass[floats, eqnum, showpacs, nofootinbib, preprint,
eqsecnum]{revtex4-1}

\usepackage{color,graphicx}
\usepackage{amsfonts}
\usepackage{amssymb}
\begin{document}

\title{Deflection angle in the strong deflection limit in a general asymptotically flat, static, spherically symmetric spacetime}
\author{Naoki Tsukamoto}\email{tsukamoto@rikkyo.ac.jp}

\affiliation{
School of Physics, Huazhong University of Science and Technology, Wuhan 430074, China
}

\begin{abstract}
Gravitational lensing by the light sphere of compact objects like black holes and wormholes will give us information on the compact objects.
In this paper, we provide an improved strong deflection limit analysis in a general asymptotically flat, static, spherically symmetric spacetime.
The strong deflection limit analysis also works in ultrastatic spacetimes.
As an example of an ultrastatic spacetime, we reexamine the deflection angle in the strong deflection limit in an Ellis wormhole spacetime. 
Using the strong deflection limit,
we obtain the deflection angle analytically for the Reissner-Nordstr\"{o}m spacetime.
The point of the improvement is the definition of a standard variable in the strong deflection limit analysis. 
We show that the choice of the variable is as important as the choice of the coordinates 
and we conclude that one should choose a proper variable for a given spacetime. 
\end{abstract}

\maketitle

\section{Introduction}
Gravitational lenses are a good tool to search the mass of dark gravitational objects between an observer and a source. 
Gravitational lenses under a weak-field approximation have been investigated 
with intensity~\cite{Schneider_Ehlers_Falco_1992,Petters_Levine_Wambsganss_2001,Schneider_Kochanek_Wambsganss_2006}
while gravitational lensing without the weak-field approximation has been also studied for a long time~\cite{Perlick_2004_Living_Rev,Bozza_2010}.
In 1959 Darwin pointed out that faint images appear near a light sphere or a photon sphere~\cite{Atkinson_1965,Claudel:2000yi,Hasse_Perlick_2002}
in the Schwarzschild spacetime~\cite{Darwin_1959}.
Images lensed by a light sphere have been revived by several authors~\cite{Atkinson_1965,Luminet_1979,Ohanian_1987,Nemiroff_1993%
,Frittelli_Kling_Newman_2000,Virbhadra_Ellis_2000,Bozza_Capozziello_Iovane_Scarpetta_2001,Bozza_2002,Perlick_2004_Phys_Rev_D%
,Iyer:2006cn,Virbhadra_Keeton_2008,Virbhadra_2009%
,Bozza_Sereno_2006,Bozza:2007gt,Bozza:2008ev%
,Eiroa:2003jf,Holz:2002uf} and
images near a light sphere of a Reissner-Nordstrom black hole~\cite{Bozza_2002,Eiroa:2002mk,Eiroa:2003jf,Tsukamoto:2016oca},
the Kerr black hole~\cite{Bozza:2002af,Bozza:2005tg,Bozza:2006nm,Saida:2016kpk},
braneworld black holes~\cite{Eiroa:2004gh,Whisker:2004gq,Eiroa:2012fb},
the other various black holes~\cite{Bhadra:2003zs,Eiroa:2005ag,Mukherjee:2006ru,Gyulchev:2006zg,Chen:2009eu,Liu:2010wh,Eiroa:2010wm,Ding:2010dc,Chen:2011ef,%
Wei:2011nj,AzregAinou:2012xv,Gyulchev:2012ty,Eiroa:2013nra,Tsukamoto:2014dta,Wei:2014dka,Eiroa:2014mca,Sahu:2015dea,Sotani:2015ewa,Zhao:2016kft,Chakraborty:2016lxo}, 
naked singularities~\cite{Virbhadra_Keeton_2008,Virbhadra_Ellis_2002,Dey_Sen_2008,Bozza_2002}, 
boson stars~\cite{Horvat:2013plm,Cunha:2015yba,Cunha:2016bjh},
and wormholes~\cite{Chetouani_Clement_1984,Nandi_Zhang_Zakharov_2006,Dey_Sen_2008,Perlick_2004_Phys_Rev_D,%
Bhattacharya:2010zzb,Gibbons_Vyska_2012,Nakajima_Asada_2012,Tsukamoto_Harada_Yajima_2012,Tsukamoto:2016zdu,Tsukamoto:2016qro,Nandi:2016ccg} 
have been investigated.

In~\cite{Bozza_2002}, Bozza presented a formalism to obtain the deflection angle $\alpha(b)$ of a light 
in the strong deflection limit $b\rightarrow b_{c}$,
where $b$ is the impact parameter of the light and $b_{c}$ is the critical impact parameter when the light ray winds around a light sphere,
in a general static spherically symmetric spacetime.
The deflection angle in the strong deflection limit $b\rightarrow b_{c}$ is expressed as
\begin{equation}
\alpha(b)=-\bar{a}\log \left( \frac{b}{b_{c}}-1 \right) +\bar{b} +O((b-b_{c})\log(b-b_{c})),
\end{equation}
where $\bar{a}$ is a positive function and $\bar{b}$ is a function.
\footnote{
The order of the error term in the deflection angle presented by Bozza~\cite{Bozza_2002} was $O(b-b_{c})$.
Recently, Tsukamoto pointed out that the error term is underestimated~\cite{Tsukamoto:2016qro}. 
}

The functions $\bar{a}$ and $\bar{b}$ in the deflection angle in the strong deflection limit are one of the fundamental values characterizing a light sphere 
or the strong-gravitational region of a spacetime.
Bozza clearly showed that the positions and the magnifications of images near a light sphere depend on $\bar{a}$ and $\bar{b}$
and the effect of $\bar{a}$ and $\bar{b}$ on the separation and the magnification of images lensed by the light sphere of
a supermassive black hole at the center of our galaxy might be measured in the near future~\cite{Bozza_2002}
and then dozens of researchers have discussed them in various spacetimes.
(See Refs.~\cite{Virbhadra_Ellis_2000,Bozza:2008ev,Virbhadra_2009} for a numerical approach.) 
Holz and Wheeler have suggested the survey of light curves of light rays which are reflected by the light sphere of a black hole near the solar system~\cite{Holz:2002uf}.
The gravitational lensing is called retrolensing and
the strong deflection limit analysis is useful to obtain retrolensing light curves~\cite{Eiroa:2003jf,Bozza:2004kq,Tsukamoto:2016qro,Tsukamoto:2016oca}.
The effect of images appearing near a light sphere on microlensing light curves also has been discussed in Refs.~\cite{Petters:2002fa,Tsukamoto:2014dta,Tsukamoto:2016qro}. 
Finite-distance corrections to the deflection angle in the strong deflection limit have been investigated in~\cite{Ishihara:2016sfv}.
Relations between the functions $\bar{a}$ and $\bar{b}$ and 
quasinormal modes~\cite{Stefanov:2010xz,Wei:2013mda} 
and high-energy absorption cross sections~\cite{Wei:2011zw} have been considered.

Recently, Tsukamoto pointed out that a strong deflection limit analysis presented by Bozza~\cite{Bozza_2002} does not work 
in ultrastatic spacetimes and obtained a deflection angle in the strong deflection limit in an ultrastatic Ellis spacetime in trial-and-error methods~\cite{Tsukamoto:2016qro}.
Here we name a spacetime with a time translational Killing vector which has a constant norm ultrastatic spacetime.

The deflection angle in a strong deflection limit for Reissner-Nordstr\"{o}m spacetime
was obtained numerically for the first time by Eiroa \textit{et al.}~\cite{Eiroa:2002mk}.
Then Bozza also calculated it numerically and claimed that $\bar{b}$ cannot be calculated analytically 
in the deflection angle with the strong deflection limit analysis provided in Ref.~\cite{Bozza_2002}.
Very recently, Tsukamoto and Gong obtained a deflection angle in the strong deflection limit analytically in Reissner-Nordstr\"{o}m spacetime~\cite{Tsukamoto:2016oca}.

In this paper, we reconsider the strong deflection limit analysis in a general asymptotically flat, static, spherically symmetric spacetime 
and provide a new formalism working well in ultrastatic spacetimes.
We show also that we can obtain $\bar{b}$ analytically in the deflection angle in the Reissner-Nordstr\"{o}m spacetime 
with the improved strong deflection limit analysis.

This paper is organized as follows. 
In Sec.~II we obtain the formula of a deflection angle in the strong deflection limit in a general asymptotically flat, static, spherically symmetric spacetime.
In Sec.~III we apply the formula to the Schwarzschild spacetime, 
the Reissner-Nordstr\"{o}m spacetime, and the Ellis wormhole spacetime.
In Sec.~IV we summarize our result.
In this paper we use the units in which the light speed and Newton's constant are unity.

\section{Deflection angle in the strong deflection limit}
In this section, we give an improved method to obtain the deflection angle of a light ray in the strong deflection limit 
in a general asymptotically flat, static, and spherically symmetric spacetime.
The line element is described by
\begin{equation}\label{eq:line_element1}
ds^{2}=-A(r)dt^{2}+B(r)dr^{2}+C(r)(d\theta^{2}+\sin^{2}\theta d\phi^{2}),
\end{equation}
where $A(r)$, $B(r)$, and $C(r)$ satisfy an asymptotically-flat condition
\footnote{
In Ref.~\cite{Bozza_2002}, the following asymptotically-flat condition is assumed:
\begin{eqnarray}\label{eq:A(r)2}
&&\lim_{r\rightarrow \infty} A(r) = 1-\frac{2M}{r},\\\label{eq:B(r)2}
&&\lim_{r\rightarrow \infty} B(r) = 1+\frac{2M}{r},\\\label{eq:C(r)2}
&&\lim_{r\rightarrow \infty} C(r) = r^{2},
\end{eqnarray}
where $M$ is the Arnowitt-Deser-Misner (ADM) mass. 
}
\begin{eqnarray}\label{eq:A(r)}
&&\lim_{r\rightarrow \infty} A(r) = 1,\\\label{eq:B(r)}
&&\lim_{r\rightarrow \infty} B(r) = 1,\\\label{eq:C(r)}
&&\lim_{r\rightarrow \infty} C(r) = r^{2}.
\end{eqnarray}

There exist time translational and axial Killing vectors $t^{\mu}\partial_{\mu}=\partial_{t}$ and $\phi^{\mu}\partial_{\mu}=\partial_{\phi}$ 
since the spacetime is static and spherical symmetric, respectively.
If a spacetime has a time translational Killing vector with a constant norm, i.e., if $A(r)$ is constant, 
we name the spacetime ultrastatic spacetime.
In any ultrastatic spacetime, we can transform constant $A(r)$ into unity without loss of generality.
Thus, ultrastatic spacetimes can satisfy the condition~(\ref{eq:A(r)}).

We assume that there is at least one positive solution of $D(r)=0$, where
\begin{equation}\label{eq:D1}
D(r)
\equiv \frac{C'(r)}{C(r)}-\frac{A'(r)}{A(r)},
\end{equation}
where $'$ denotes the differentiation with respect to the radial coordinate $r$. 
We call the largest positive solution of $D(r)=0$ the radius of a light sphere $r_{m}$.
We assume that $A(r)$, $B(r)$, and $C(r)$ are finite and positive for $r\geq r_{m}$.
\footnote{
In Ref.~\cite{Bozza_2002}, $A'(r)>0$ and $C'(r)>0$ for $r>r_{m}$ are also assumed. 
We extend a formalism presented by Bozza~\cite{Bozza_2002} to obtain 
a deflection angle in the strong deflection limit in ultrastatic spacetimes with $A'(r)=0$ everywhere.
}

The trajectory of a light is described by $g_{\mu \nu}k^{\mu}k^{\nu}=0$, where $k^{\mu}\equiv \dot{x}^{\mu}$ is the wave number of the photon 
and $\dot{\:}$ denotes the differentiation with respect to an affine parameter parametrizing the trajectory. 
The conserved energy $E\equiv -g_{\mu\nu}t^{\mu}k^{\nu}=A(r)\dot{t}$ 
and the conserved angular momentum $L\equiv g_{\mu\nu}\phi^{\mu}k^{\nu}=C(r)\dot{\phi}$ are constant along it.
We assume that $E$ and $L$ do not vanish.
We define the impact parameter $b$ as 
\begin{equation}\label{eq:impact_parameter1}
b\equiv \frac{L}{E} 
=\frac{C(r)\dot{\phi}}{A(r)\dot{t}}.
\end{equation}
Without loss of generality, we can assume $\theta=\pi/2$ because of spherical symmetry.
The trajectory equation is expressed as
\begin{equation}\label{eq:trajectory1}
-A(r)\dot{t}^{2}+B(r)\dot{r}^{2}+C(r)\dot{\phi}^{2}=0
\end{equation}
or
\begin{equation}\label{eq:trajectory2}
\dot{r}^{2}=V(r),
\end{equation}
where the effective potential $V(r)$ for the motion of a photon is defined as
\begin{equation}\label{eq:potential}
V(r)\equiv \frac{L^{2}R(r)}{B(r)C(r)},
\end{equation}
where $R(r)$ is given by
\begin{equation}\label{eq:R1}
R(r) \equiv \frac{C(r)}{A(r)b^{2}}-1.
\end{equation}
The motion of the photon is permitted in the region $V(r)\geq 0$.
Since we obtain $\lim_{r \rightarrow \infty}V(r)=E^{2}>0$ from the asymptotically-flat condition~(\ref{eq:A(r)})-(\ref{eq:C(r)}), 
the photon can exist at infinity $r\rightarrow \infty$.
We assume that $R(r)=0$ has at least one positive solution. 

We consider that a photon approaches a gravitational object from infinity, is scattered at a closest distance $r=r_{0}$, and goes to infinity.
In the scatter case, $r_{m}<r_{0}$ should be satisfied.
Please note that $r=r_{0}$ is the largest positive solution of $R(r)=0$
and that $B(r)$ and $C(r)$ do not diverge for $r\geq r_{0}$.
Thus, $V(r)$ vanishes at $r=r_{0}$.
Since $\dot{r}$ vanishes at the closest distance $r=r_{0}$, from the trajectory equation~(\ref{eq:trajectory1}), we obtain
\begin{equation}\label{eq:at_closest}
A_{0}\dot{t}^{2}_{0}=C_{0}\dot{\phi}^{2}_{0}.
\end{equation}
Here and hereafter subscript $0$ denotes the quantities at $r=r_{0}$.
Without loss of generality, we can assume that the impact parameter $b$ is positive as long as we consider only one light ray.
Since the impact parameter is constant along the trajectory, 
using Eq.~(\ref{eq:at_closest}), the impact parameter~(\ref{eq:impact_parameter1}) can be expressed as 
\begin{equation}\label{eq:impact_parameter2}
b(r_{0})=\frac{L}{E}=\frac{C_{0}\dot{\phi}_{0}}{A_{0}\dot{t}_{0}}=\sqrt{\frac{C_{0}}{A_{0}}}. 
\end{equation}
Using Eq.~(\ref{eq:impact_parameter2}), we can rewrite $R(r)$ into
\begin{equation}\label{eq:R2}
R(r)= \frac{A_{0}C}{AC_{0}}-1.
\end{equation}

We show a necessary and sufficient condition for the existence of a circular light orbit by following Hasse and Perlick~\cite{Hasse_Perlick_2002}.
We express the trajectory equation as 
\begin{equation}\label{eq:trajectory4}
\frac{BC\dot{r}^{2}}{E^{2}}+b^{2}=\frac{C}{A}.
\end{equation}
Differentiating Eq.~(\ref{eq:trajectory4}) with respect to the affine parameter and then dividing it by $\dot{r}$,
we obtain 
\begin{equation}\label{eq:trajectory5}
\ddot{r}+\frac{1}{2}\left( \frac{B'}{B}+\frac{C'}{C} \right) \dot{r}^{2}=\frac{E^{2}}{AB}D(r).
\end{equation}
Since $A(r)$, $B(r)$ ,and $C(r)$ are finite and positive for $r\geq r_{m}$ and $E$ is positive, a circular light orbit exists if and only if 
\begin{equation}\label{eq:light_sphere1}
D(r)=0.
\end{equation}
Please note that $R'_{m}=D_{m}C_{m}A_{m}/b^{2}=0$, where subscript $m$ denotes the quantities at $r=r_{m}$.

We define the critical impact parameter $b_{c}$ as 
\begin{equation}\label{eq:critical_impact_parameter1}
b_{c}(r_{m})
\equiv \lim_{r_{0}\rightarrow r_{m}} \sqrt{\frac{C_{0}}{A_{0}}},
\end{equation}
and name a limit $r_{0}\rightarrow r_{m}$ or $b\rightarrow b_{c}$ strong deflection limit. 
The derivative of the effective potential $V(r)$ with respect to $r$ is given by
\begin{equation}
V'(r)=\frac{L^{2}}{BC}\left[ R'+\left(\frac{C'}{C}-\frac{B'}{B}\right)R \right],
\end{equation}
and, hence, we obtain 
\begin{equation}\label{eq:potential_critical1}
\lim_{r_{0}\rightarrow r_{m}} V(r_{0})
=\lim_{r_{0}\rightarrow r_{m}} V'(r_{0})
=0
\end{equation}
in the strong deflection limit $r_{0}\rightarrow r_{m}$.
This means that the light ray winds around the light sphere in the strong deflection limit.

The trajectory equation of a light is rewritten as
\begin{equation}\label{eq:trajectory3}
\left(\frac{dr}{d\phi}\right)^{2}=\frac{R(r)C(r)}{B(r)},
\end{equation}
and the deflection angle $\alpha(r_{0})$ of the light is obtained as
\begin{equation}\label{eq:deflection1}
\alpha(r_{0})=I(r_{0})-\pi,
\end{equation}
where $I(r_{0})$ is defined as
\begin{equation}\label{eq:I1}
I(r_{0})\equiv 2\int^{\infty}_{r_{0}}\frac{dr}{\sqrt{\frac{R(r)C(r)}{B(r)}}}.
\end{equation}

Introducing a variable $z$ defined as 
\footnote{
In Ref.~\cite{Bozza_2002}, a counterpart $z_{[16]}$ of the variable $z$ is defined as 
\begin{equation}\label{eq:z1}
z_{[16]}\equiv \frac{A-A_{0}}{1-A_{0}}.
\end{equation}
We discuss the details of $z$ and $z_{[16]}$ in Secs.~III and IV.
}
\begin{equation}\label{eq:z}
z\equiv 1-\frac{r_{0}}{r},
\end{equation}
$I(r_{0})$ is described by
\begin{equation}\label{eq:I2}
I(r_{0})=\int^{1}_{0}f(z,r_{0})dz,
\end{equation}
where $f(z,r_{0})$ is defined as
\begin{equation}\label{eq:f1}
f(z,r_{0})
\equiv \frac{2r_{0}}{\sqrt{G(z,r_{0})}},
\end{equation}
where $G(z,r_{0})$ is given by
\begin{equation}\label{eq:G1}
G(z,r_{0})\equiv R\frac{C}{B}(1-z)^{4}.
\end{equation}

Since the expansions of a function $F(r)$ and its inverse $1/F(r)$ in the power of $z$ are given by
\begin{equation}\label{eq:expansion_F}
F=F_{0}+F^{'}_{0}r_{0}z+\left( \frac{1}{2}F^{''}_{0}r^{2}_{0}+F^{'}_{0}r_{0} \right)z^{2}+O(z^{3})
\end{equation}
and
\begin{eqnarray}\label{eq:expansion_1/F}
\frac{1}{F}
&=&\frac{1}{F_{0}}-\frac{F_{0}^{'}r_{0}}{F_{0}^{2}}z\nonumber\\
&&+\frac{r_{0}}{F_{0}^{3}} \left( -\frac{1}{2}F_{0}F_{0}^{''}r_{0} +F_{0}^{'2}r_{0}-F_{0}F_{0}^{'} \right)z^{2} +O(z^{3}),\nonumber\\
\end{eqnarray}
respectively,
$R(r)$ can be expanded in the power of $z$ as
\begin{eqnarray}\label{eq:R3}
R(r)
&=&D_{0}r_{0}z 
+\left[ \frac{r_{0}}{2} \left( \frac{C_{0}^{''}}{C_{0}}-\frac{A_{0}^{''}}{A_{0}} \right) \right. \nonumber\\
&&+\left. \left( 1-\frac{A^{'}_{0}r_{0}}{A_{0}} \right) D_{0} \right] r_{0}z^{2} +O(z^{3}). 
\end{eqnarray}
Using Eqs.~(\ref{eq:G1})-(\ref{eq:R3}),  we obtain the expansion of $G(z,r_{0})$ in the power of $z$ as 
\begin{equation}\label{eq:G2}
G(z,r_{0})=\sum^{\infty}_{n=1} c_{n}(r_{0})z^{n},
\end{equation}
where $c_{1}(r_{0})$ and $c_{2}(r_{0})$ are given by
\begin{equation}\label{eq:c10}
c_{1}(r_{0})=\frac{C_{0}D_{0}r_{0}}{B_{0}}
\end{equation}
and
\begin{eqnarray}\label{eq:c20}
c_{2}(r_{0})
&=&\frac{C_{0}r_{0}}{B_{0}} \left\{ D_{0} \left[  \left( D_{0}-\frac{B^{'}_{0}}{B_{0}} \right)r_{0}-3 \right] \right. \nonumber\\
&&\left. +\frac{r_{0}}{2} \left( \frac{C^{''}_{0}}{C_{0}}-\frac{A^{''}_{0}}{A_{0}} \right) \right\},
\end{eqnarray}
respectively.

In the strong deflection limit $r_{0}\rightarrow r_{m}$, from $D_{m}=0$, we obtain
\begin{equation}\label{eq:c1m}
c_{1}(r_{m})=0
\end{equation}
and
\begin{equation}\label{eq:c2m}
c_{2}(r_{m})=\frac{C_{m}r_{m}^{2}}{2B_{m}} D^{'}_{m} ,
\end{equation}
where 
\begin{equation}\label{eq:D'm}
D^{'}_{m} = \frac{C_{m}^{''}}{C_{m}}-\frac{A_{m}^{''}}{A_{m}} 
\end{equation}
and, hence, we obtain
\begin{equation}\label{eq:G3}
G_{m}(z)=c_{2}(r_{m})z^{2}+O(z^{3}).
\end{equation}
This shows that the leading order of the divergence of $f(z, r_{0})$ is $z^{-1}$ 
and that the integral $I(r_{0})$ diverges logarithmically in the strong deflection limit~$r_{0}\rightarrow r_{m}$.

We separate the integral $I(r_{0})$ into a divergent part $I_{D}(r_{0})$ and a regular part $I_{R}(r_{0})$.
We define the divergent part $I_{D}(r_{0})$ as 
\begin{equation}\label{eq:ID1}
I_{D}(r_{0})
\equiv \int^{1}_{0}f_{D}(z,r_{0})dz,
\end{equation}
where $f_{D}(z,r_{0})$ is defined by
\begin{equation}\label{eq:fD1}
f_{D}(z,r_{0})
\equiv \frac{2r_{0}}{\sqrt{c_{1}(r_{0})z+c_{2}(r_{0})z^{2}}}.
\end{equation}
We can integrate $I_{D}(r_{0})$ and obtain 
\begin{equation}\label{eq:ID2}
I_{D}(r_{0})=\frac{4r_{0}}{\sqrt{c_{2}(r_{0})}}\log \frac{\sqrt{c_{2}(r_{0})}+\sqrt{c_{1}(r_{0})+c_{2}(r_{0})}}{\sqrt{c_{1}(r_{0})}}.
\end{equation}

Since the expansions of $c_{1}(r_{0})$ and $b(r_{0})$ in powers of $r_{0}-r_{m}$ are given by
\begin{equation}\label{eq:c1_expansion}
c_{1}(r_{0})=\frac{C_{m}r_{m}D^{'}_{m}}{B_{m}}(r_{0}-r_{m})+O((r_{0}-r_{m})^{2})
\end{equation}
and 
\begin{equation}\label{eq:b_expansion}
b(r_{0})=b_{c}(r_{m})+\frac{1}{4}\sqrt{\frac{C_{m}}{A_{m}}}D^{'}_{m}(r_{0}-r_{m})^{2}+O((r_{0}-r_{m})^{3}),
\end{equation}
respectively,
$c_{1}(r_{0})$ in the strong deflection limit $r_{0}\rightarrow r_{m}$ is described by 
\begin{equation}\label{eq:cm1}
\lim_{r_{0}\rightarrow r_{m}} c_{1}(r_{0}) 
= \lim_{b\rightarrow b_{c}} \frac{2C_{m}r_{m}\sqrt{D^{'}_{m}}}{B_{m}} \left( \frac{b}{b_{c}}-1 \right)^{\frac{1}{2}}.
\end{equation}
Thus, we obtain the divergent part $I_{D}(b)$ in the strong deflection limit $b\rightarrow b_{c}$ as 
\begin{eqnarray}\label{eq:IDm1}
I_{D}(b)
&=&-{\frac{r_{m}}{\sqrt{c_{2}(r_{m})}}} \log \left( \frac{b}{b_{c}}-1 \right) +{\frac{r_{m}}{\sqrt{c_{2}(r_{m})}}} \log r^{2}_{m}D^{'}_{m}\nonumber\\
&&+O((b-b_{c})\log (b-b_{c})).
\end{eqnarray}

We define the regular part $I_{R}(r_{0})$ as
\begin{equation}\label{eq:IR1}
I_{R}(r_{0})
\equiv \int^{1}_{0} f_{R}(z,r_{0})dz,
\end{equation}
where $f_{R}(r_{0})$ is defined by
\begin{equation}\label{eq:fR1}
f_{R}(r_{0})
\equiv f(z,r_{0})-f_{D}(z,r_{0}).
\end{equation}
Please notice $I(r_{0})=I_{D}(r_{0})+I_{R}(r_{0})$.
We expand $I_{R}(r_{0})$ in powers of $r_{0}-r_{m}$ and we concentrate on the leading term $I_{R}(r_{m})$
since we are interested in the regular part $I_{R}$ in the strong deflection limit $r_{0}\rightarrow r_{m}$ or $b\rightarrow b_{c}$.
We should integrate analytically or numerically the regular part
\begin{equation}\label{eq:IR2}
I_{R}(r_{0})
= \int^{1}_{0} f_{R}(z,r_{m})dz+O((r_{0}-r_{m})\log(r_{0}-r_{m}))
\end{equation}
or 
\begin{equation}\label{eq:IR3}
I_{R}(b)
= \int^{1}_{0} f_{R}(z,b_{c})dz+O((b-b_{c})\log(b-b_{c})).
\end{equation}

The deflection angle in the strong deflection limit $r_{0}\rightarrow r_{m}$ or $b\rightarrow b_{c}$ is given by
\begin{equation}\label{eq:alpha2}
\alpha(b)=-\bar{a}\log \left( \frac{b}{b_{c}}-1 \right) +\bar{b} +O((b-b_{c})\log(b-b_{c})),
\end{equation}
where $\bar{a}$ and $\bar{b}$ are given by
\begin{equation}\label{eq:abar1}
\bar{a}=\sqrt{\frac{2B_{m}A_{m}}{C^{''}_{m}A_{m}-C_{m}A^{''}_{m}}}
\end{equation}
and
\begin{equation}\label{eq:bbar1}
\bar{b}=\bar{a}\log \left[r^{2}_{m}\left(\frac{C_{m}^{''}}{C_{m}}-\frac{A_{m}^{''}}{A_{m}}\right)\right] +I_{R}(r_{m})-\pi,
\end{equation}
respectively.

\section{Applications}
In this section, we apply our result obtained in the previous section for the Schwarzschild spacetime, 
the Reissner-Nordstr\"om spacetime, and an Ellis wormhole spacetime.

\subsection{Schwarzschild black hole}
The line element in the Schwarzschild spacetime is given by 
\begin{equation}\label{eq:S_metric}
ds^{2}=-\left(1-\frac{2M}{r} \right)dt^{2}+\frac{dr^{2}}{1-\frac{2M}{r}}+r^{2}(d\theta^{2}+\sin^{2}\theta d\phi^{2})
\end{equation}
and then
\begin{eqnarray}
&&A(r)=1-\frac{2M}{r},\\
&&B(r)=\frac{1}{1-\frac{2M}{r}},\\
&&C(r)=r^{2}.
\end{eqnarray}
From Eq.~(\ref{eq:D1}), 
we obtain $r_{m}=3M$.
There is an event horizon at $r=r_{H}\equiv2M$ and $B(r)$ diverges there 
but all the assumptions are satisfied 
because of $r_{m}>r_{H}$. 

From Eq.~(\ref{eq:critical_impact_parameter1}), the critical impact parameter $b_{c}$ is given by
\begin{equation}
b_{c}=3\sqrt{3}M.
\end{equation}
From Eqs.~(\ref{eq:abar1}) and (\ref{eq:bbar1}) we obtain 
\begin{equation}
\bar{a}=1
\end{equation}
and 
\begin{equation}
\bar{b}=\log 6+I_{R}(r_{m})-\pi,
\end{equation}
respectively.

From Eq.~(\ref{eq:IR2}), we obtain $I_{R}(r_{m})$ as
\begin{eqnarray}
I_{R}(r_{m})
&=&2\int^{1}_{0} \left( \frac{1}{z\sqrt{1-\frac{2}{3}z}} -\frac{1}{z} \right) dz \nonumber\\
&=&2\log [6(2-\sqrt{3})],
\end{eqnarray}
and, hence, we obtain $\bar{b}$ as
\begin{equation}
\bar{b}=\log \lambda -\pi,
\end{equation}
where $\lambda\equiv 216(7-4\sqrt{3})$.
This is the same result obtained in Ref.~\cite{Bozza_2002}.

In Ref.~\cite{Bozza_2002}, Bozza used a counterpart $z_{[16]}$ of the variable $z$ defined as 
\begin{equation}\label{eq:z3}
z_{[16]}\equiv \frac{A-A_{0}}{1-A_{0}}.
\end{equation}
We realize that the formalism presented in Sec~II to obtain the deflection angle in the strong deflection angle 
is the same as the one in \cite{Bozza_2002} in the Schwarzschild spacetime
since $z$ is equivalent to $z_{[16]}$:  
\begin{equation}
z=z_{[16]}=1-\frac{r_{0}}{r}.
\end{equation}

Iyer and Petters~\cite{Iyer:2006cn} expanded the deflection angle in the Schwarzschild spacetime as an affine perturbation series 
\begin{eqnarray}
\alpha 
&=&-\log \left( \frac{b}{b_{c}}-1 \right) +\log \lambda -\pi \nonumber\\
&&+\sum^{\infty}_{n=1} \left( \frac{b}{b_{c}}-1 \right)^{n} \left\{ \rho_{n}-\sigma_{n} \log \left[ \left( \frac{b}{b_{c}}-1 \right) \frac{1}{\lambda} \right]  \right\}, \nonumber\\
\end{eqnarray}
where $\rho_{n}$ and $\sigma_{n}$ are constant.
They showed $\rho_{n}$ and $\sigma_{n}$ clearly for $n \leq 3$.
This fact shows that the order of the error term of the deflection angle in the strong deflection limit is not $O(b_{c}-b)$ but $O((b_{c}-b)\log (b_{c}-b))$.

\subsection{Reissner-Nordstr\"om black hole}
Very recently, Tsukamoto and Gong obtained the deflection angle in the strong deflection limit analytically 
in the Reissner-Nordstr\"om spacetime~\cite{Tsukamoto:2016oca}.
They used the variable $z$ defined as Eq.~(\ref{eq:z}). 
We apply the formula of the deflection angle in the strong deflection limit presented in Section~II 
for the Reissner-Nordstr\"om spacetime to test the formula.

The line element of the Reissner-Nordstr\"om spacetime is given by
\begin{eqnarray}\label{eq:RN_metric}
ds^{2}&=&-\left(1-\frac{2M}{r}+\frac{Q^{2}}{r^{2}} \right)dt^{2}+\frac{dr^{2}}{1-\frac{2M}{r}+\frac{Q^{2}}{r^{2}}}\nonumber\\
&&+r^{2}(d\theta^{2}+\sin^{2}\theta d\phi^{2})
\end{eqnarray}
and $A(r)$, $B(r)$, and $C(r)$ are given by
\begin{eqnarray}
&&A(r)=1-\frac{2M}{r}+\frac{Q^{2}}{r^{2}},\\
&&B(r)=\frac{1}{1-\frac{2M}{r}+\frac{Q^{2}}{r^{2}}},\\
&&C(r)=r^{2}.
\end{eqnarray}
If the electrical charge $Q$ satisfies $Q\leq M$, there is an event horizon at $r=r_{H}\equiv M+\sqrt{M^{2}-Q^{2}}$. 
We concentrate on the black hole spacetime with $Q\leq M$. 
From Eq.~(\ref{eq:D1}), 
There is a light sphere at $r=r_{m}$, 
where $r_{m}$ is given by
\begin{equation}\label{eq:rm_q}
r_{m}=\frac{3M+\sqrt{9M^{2}-8Q^{2}}}{2}.
\end{equation}
Note that $B(r)$ is finite and positive in the range $r \geq r_{m}$ since $r_{H}<r_{m}$.

From Eq.~(\ref{eq:critical_impact_parameter1}), the critical impact parameter $b_{c}$ is given by
\begin{equation}
b_{c}=\frac{r_{m}^{2}}{\sqrt{Mr_{m}-Q^{2}}}.
\end{equation}
We obtain $\bar{a}$ and $\bar{b}$ as, from Eqs.~(\ref{eq:abar1}) and (\ref{eq:bbar1}), 
\begin{equation}\label{eq:bara_q}
\bar{a}=\frac{r_{m}}{\sqrt{3Mr_{m}-4Q^{2}}}
\end{equation}
and
\begin{equation}
\bar{b}=\bar{a} \log \frac{2(3Mr_{m}-4Q^{2})}{Mr_{m}-Q^{2}}+I_{R}(r_{m})-\pi,
\end{equation}
respectively.
From Eq.~(\ref{eq:IR2}), the regular integral is given by
\begin{eqnarray}
I_{R}(r_{m})
&=&\int^{1}_{0}  \frac{2r_{m}dz}{z\sqrt{3Mr_{m}-4Q^{2}-2(Mr_{m}-2Q^{2})z-Q^{2}z^{2}}}  \nonumber\\
&&-\int^{1}_{0} \frac{2\bar{a}}{z} dz \nonumber\\
&=&\bar{a} \log \left[ \frac{4(3Mr_{m}-4Q^{2})^{2}}{M^{2}r_{m}^{2}(Mr_{m}-Q^{2})} \right. \nonumber\\
&& \left. \times \left(2\sqrt{Mr_{m}-Q^{2}}-\sqrt{3Mr_{m}-4Q^{2}}\right)^{2}  \right]
\end{eqnarray}
Thus, we obtain $\bar{b}$ as
\begin{eqnarray}\label{eq:barb_q}
\bar{b}
&=&\bar{a} \log \left[ \frac{8(3Mr_{m}-4Q^{2})^{3}}{M^{2}r_{m}^{2}(Mr_{m}-Q^{2})^{2}} \right. \nonumber\\
&&\left. \times \left(2\sqrt{Mr_{m}-Q^{2}}-\sqrt{3Mr_{m}-4Q^{2}}\right)^{2} \right]   -\pi. \nonumber\\
\end{eqnarray}
When $Q=0$, we obtain $b_{c}$, $\bar{a}$, and $\bar{b}$ which are equal to $b_{c}$, $\bar{a}$, and $\bar{b}$ in the Schwarzschild spacetime.
Then the maximal charged black hole case $(Q=M)$, 
$b_{c}$, $\bar{a}$, and $\bar{b}$ becomes
\begin{equation}
b_{c}=4M,
\end{equation}
\begin{equation}
\bar{a}=\sqrt{2},
\end{equation}
and 
\begin{equation}
\bar{b}=2\sqrt{2}\log [4(2-\sqrt{2})]-\pi,
\end{equation}
respectively.
Figure~1 shows $b_{c}/M$, $\bar{a}$, and $\bar{b}$ as a function of $Q/M$.
\begin{figure}[htbp]
\begin{center}
\includegraphics[width=70mm]{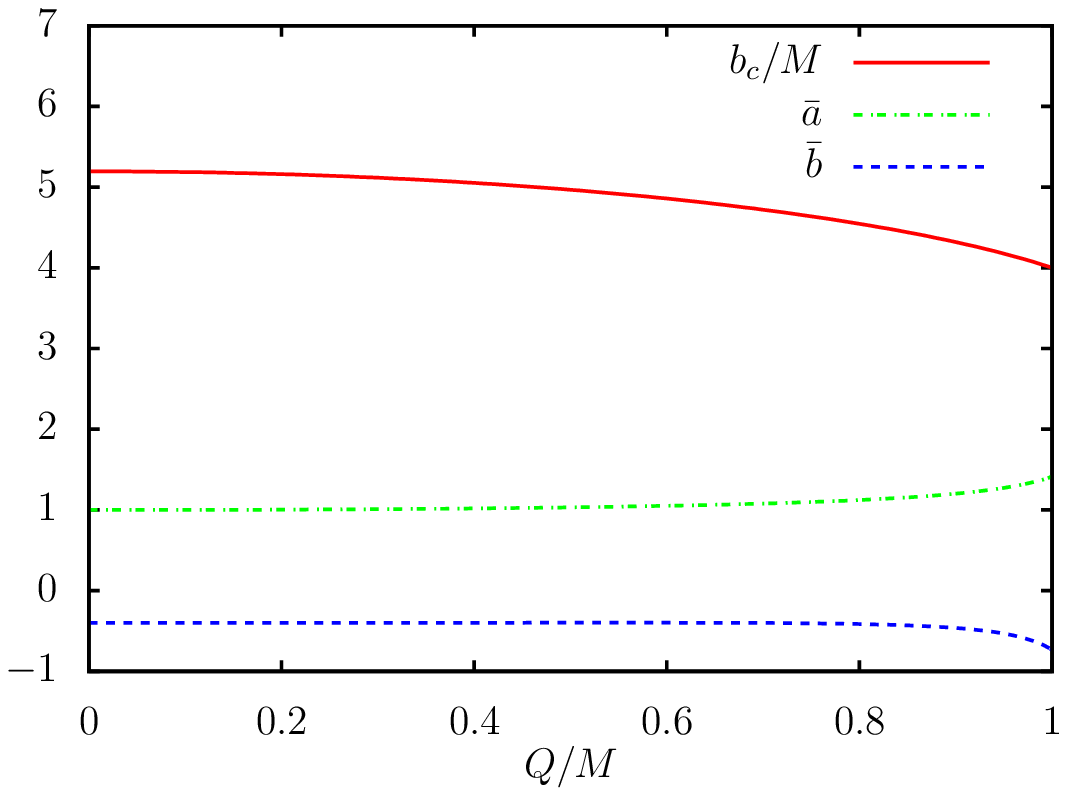}
\end{center}
\caption{$b_{c}/M$, $\bar{a}$, and $\bar{b}$ in the Reissner-Nordstr\"om black hole spacetime.
The solid (red), dot-dashed (green),
and dashed (blue) curves denote $b_{c}/M$, $\bar{a}$, and $\bar{b}$, respectively.}
\end{figure}
These are the same ones obtained in Ref.~\cite{Tsukamoto:2016oca}.

In Ref.~\cite{Bozza_2002}, Bozza calculated numerically the counterpart of the regular integral $I_{R}(r_{m})$
but could not calculate it analytically without expanding it in powers of $(Q/M)^{2}$.
We notice that the variable $z_{[16]}$ in the Reissner-Nordstr\"om spacetime becomes
\begin{equation}\label{eq:z4}
z_{[16]}=1-\frac{r_{0}^{2}(2Mr-Q^{2})}{r^{2}(2Mr_{0}-Q^{2})}.
\end{equation}
The variable $z$ is better than $z_{[16]}$ in the Reissner-Nordstr\"om spacetime 
because it allows a complete analytical treatment.

In Ref.~\cite{Eiroa:2002mk} the deflection angle in the strong deflection limit $r_{0}\rightarrow r_{m}$ in the Reissner-Nordstr\"{o}m spacetime
was obtained numerically.
The deflection angle in the strong deflection limit $r_{0}\rightarrow r_{m}$ can be expressed as, in our notation, 
\begin{equation}
\alpha =- \mathcal{A} \log \left[ \frac{\mathcal{B}(r_{0}-r_{m})}{2M} \right] -\pi,
\end{equation}
where
\begin{equation}
\mathcal{A}\equiv 2\bar{a},
\end{equation}
\begin{equation}
\mathcal{B}\equiv M\sqrt{D_{m}^{'}}\exp \left( -\frac{\bar{b}+\pi}{2\bar{a}} \right),
\end{equation}
and 
\begin{equation}
D_{m}^{'}=\frac{2(3Mr_{m}-4Q^{2})}{r^{2}_{m}(Mr_{m}-Q^{2})}.
\end{equation}
We can obtain the analytic formula of $\mathcal{A}$ and $\mathcal{B}$ from Eqs.~(\ref{eq:rm_q}), (\ref{eq:bara_q}), and (\ref{eq:barb_q}).
Table~\ref{table:I} shows $\mathcal{A}$ and $\mathcal{B}$ as functions of $Q/M$.
We notice that $\mathcal{A}$ and $\mathcal{B}$ with our analytic formula reproduce Table 1 of Ref. [26] with high precision.

\begin{table}[hbtp]
 \caption{$\mathcal{A}$ and $\mathcal{B}$ in the deflection angle in the strong deflection limit. One sees the same table in Ref.~\cite{Tsukamoto:2016oca}.}
 \label{table:I}
\begin{center}
\begin{tabular}{c c c c c c c} \hline
$Q/M$ &$0$ &$0.1$ &$0.25$ &$0.5$ &$0.75$ &$1$ \\ \hline
$\mathcal{A}$ &$2.00000$  &$2.00224$  &$2.01444$  &$2.06586$  &$2.19737$  &$2.82843$ \\ \hline
$\mathcal{A}$ in~\cite{Eiroa:2002mk} &$2.00000$  &$2.00224$  &$2.01444$  &$2.06586$  &$2.19737$  &$2.82843$ \\ \hline
$\mathcal{B}$ &$0.207336$ &$0.207977$ &$0.211467$ &$0.225996$ &$0.262083$ &$0.426777$ \\ \hline
$\mathcal{B}$ in~\cite{Eiroa:2002mk} &$0.207338$ &$0.207979$ &$0.21147$  &$0.225997$ &$0.262085$ &$0.426782$ \\ \hline
\end{tabular}
\end{center}
\end{table}

\subsection{Ellis wormhole}
An Ellis wormhole~\cite{Ellis_1973,Bronnikov_1973} is one of the simplest Morris-Thorne wormhole solutions~\cite{Morris_Thorne_1988}.
Gravitational lensing by the Ellis wormhole was investigated~\cite{Chetouani_Clement_1984,Perlick_2004_Phys_Rev_D,Nandi_Zhang_Zakharov_2006,Dey_Sen_2008,Muller:2008zza,Abe_2010,Toki_Kitamura_Asada_Abe_2011,Bhattacharya:2010zzb,Gibbons_Vyska_2012,Nakajima_Asada_2012,Tsukamoto_Harada_Yajima_2012,Tsukamoto_Harada_2013,Kitamura_Nakajima_Asada_2013,Izumi_Hagiwara_Nakajima_Kitamura_Asada_2013,Kitamura_Izumi_Nakajima_Hagiwara_Asada_2013,Nakajima:2014nba,Bozza:2015haa,Bozza:2015wbw,Lukmanova_2016,Takahashi_Asada_2013,Yoo_Harada_Tsukamoto_2013}
and the upper bound of the number density~\cite{Takahashi_Asada_2013,Yoo_Harada_Tsukamoto_2013} was given from surveys of gravitational lensing.
The line element in the Ellis wormhole spacetime is given by
\begin{equation}\label{eq:E_metric}
ds^{2}=-dt^{2}+dl^{2}+(l^{2}+a^{2})(d\theta^{2}+\sin^{2}\theta d\phi^{2}),
\end{equation}
where $l$ is a radial coordinate defined in the range $-\infty<l<\infty$ and $a$ is a positive constant.
There is a throat at $l=0$.
We concentrate on a light ray which does not pass the throat and which exists in a region $l\geq  0$. 
The Ellis wormhole has vanishing ADM masses~\cite{Tsukamoto:2016qro}. 
It is an ultrastatic spacetime since there is a time translational Killing vector with a constant norm. 
By solving an equation
\begin{equation}
g_{tt}(l)\frac{dg_{\theta\theta}(l)}{dl}-g_{\theta\theta}(l)\frac{dg_{tt}(l)}{dl}=0,
\end{equation}
we find a light sphere at $l=0$.
Since we have assumed a positive radius of a light sphere in the previous section, 
we introduce a new radius coordinate $r\equiv l+p$, where $p$ is a positive constant.
The line element is rewritten as 
\begin{equation}\label{eq:E_metric2}
ds^{2}=-dt^{2}+dr^{2}+[(r-p)^{2}+a^{2}](d\theta^{2}+\sin^{2}\theta d\phi^{2})
\end{equation}
and we obtain
\begin{eqnarray}
&&A(r)=1,\\
&&B(r)=1,\\
&&C(r)=(r-p)^{2}+a^{2}.
\end{eqnarray}
Under the radial coordinate $r$, the radius of the light sphere $r_{m}=p$ is positive.
The light sphere at $r=r_{m}$ is coincident with the wormhole throat. 
Please note that one cannot apply the strong deflection limit analysis in Ref.~\cite{Bozza_2002} for any ultrastatic spacetime 
because of the violation of an assumption that $A'(r)$ is positive for $r>r_{m}$
while one can apply our result investigated in Sec.~II for the Ellis wormhole spacetime with the line element~(\ref{eq:E_metric2}).
In general, we cannot define the variable $z_{[16]}$ in an ultrastatic spacetime satisfying the asymptotically-flat condition~(\ref{eq:A(r)}).

From Eq.~(\ref{eq:critical_impact_parameter1}), we obtain the critical impact parameter $b_{c}$ as 
\begin{equation}
b_{c}=a.
\end{equation}
Using Eqs.~(\ref{eq:abar1}) and (\ref{eq:bbar1}), we obtain $\bar{a}$ and $\bar{b}$ as
\begin{equation}
\bar{a}=1
\end{equation}
and
\begin{equation}
\bar{b}=\log \frac{2p^{2}}{a^{2}} +I_{R}(r_{m})-\pi,
\end{equation}
respectively.
From Eq.~(\ref{eq:IR2}), we can obtain the regular integral $I_{R}$ as
\begin{eqnarray}
I_{R}(r_{m})
&=&2\int^{1}_{0} \left( \frac{a}{z\sqrt{a^{2}-2a^{2}z+(a^{2}+p^{2})z^{2}}}- \frac{1}{z} \right) dz \nonumber\\
&=&2\log \frac{2a}{p},
\end{eqnarray}
and hence we obtain
\begin{equation}
\bar{b}=3\log 2-\pi.
\end{equation}

The deflection angle is equal to the deflection angle in the strong deflection limit obtained in Ref.~\cite{Tsukamoto:2016qro}.
Tsukamoto~\cite{Tsukamoto:2016qro} used a variable $z_{[68]}$ defined as
\begin{equation}
z_{[68]}
\equiv 1-\frac{b}{\sqrt{r^{2}+a^{2}}}=1-\frac{\sqrt{r^{2}_{0}+a^{2}}}{\sqrt{r^{2}+a^{2}}}.
\end{equation}
Since $z_{[68]}$ is not equivalent to $z$ defined as Eq.~(\ref{eq:z}) and $z_{[16]}$ in the Ellis wormhole spacetime, 
the calculation which we have presented here can be an additional cross-check 
for the deflection angle in the strong deflection limit in the Ellis wormhole spacetime.
See~\cite{Tsukamoto:2016qro} for the details of the deflection angle in the Ellis wormhole spacetime.

\section{Conclusion}
In this paper, we have investigated a strong deflection limit analysis in a general asymptotically flat, static, spherically symmetric spacetime.
The improved strong deflection limit analysis works in ultrastatic spacetimes while the one presented in Ref.~\cite{Bozza_2002} does not.
As an example of an ultrastatic spacetime, we have investigated the deflection angle in the strong deflection limit in an Ellis wormhole spacetime. 
Using the improved strong deflection limit analysis, we have obtained the deflection angle in the strong deflection limit analytically 
in the Reissner-Nordstr\"{o}m spacetime while it cannot obtained in Ref.~\cite{Bozza_2002}.
The most important point of the improvement of the strong deflection limit analysis is the definition of a variable $z$ defined as Eq.~(\ref{eq:z}). 
In this paper, we have chosen a simple variable $z$ not only in the Schwarzschild spacetime but also in the other spacetimes.
We do not insist on that we have chosen the best definition of the variable $z$ in this paper
but we show clearly that the choice of the variable $z$ is as important as the choice of the coordinates.
We conclude that one should choose a proper variable $z$ for a given spacetime. 
Even if the variable $z$ defined as Eq.~(\ref{eq:z}) is not suitable in a specific spacetime, 
our strong deflection limit analysis in this paper will give us a clue to find a more proper variable $z$ in the spacetime.

\section*{Acknowledgements}
The author thanks Ken-ichi Nakao, Tetsuya Shiromizu, Chul-Moon Yoo, Takahisa Igata, and Yungui Gong for valuable comments.
He also thanks Tomohiro Harada, Yoshimune Tomikawa, Hideki Asada, Hirotaka Yoshino, Yusuke Suzuki, Rio Saitou, Masato Nozawa, and Takafumi Kokubu for useful conversations. 
He are grateful to Carlos~A.~R.~Herdeiro for bringing his attention to gravitational lensing by the photon surface of boson stars.
This research was supported in part by the National Natural Science Foundation of China under Grant No. 11475065.
%

\end{document}